# Ecosystems perspective on financial networks: diagnostic tools

Eduardo Viegas[†]     Misako Takayasu[††]

Wataru Miura[††]     Koutarou Tamura[††]     Takaaki Ohnishi[§]

Hideki Takayasu[‡‡ §§]     Henrik Jeldtoft Jensen[†]

***Abstract:*** *The economical world consists of a highly interconnected and interdependent network of firms[1][2][3]. Here we develop temporal and structural network tools to analyze the state of the economy. Our analysis indicates that a strong clustering can be a warning sign. Reduction in diversity, which was an essential aspect of the dynamics surrounding the crash in 2008, is seen as a key emergent feature arising naturally from the evolutionary and adaptive dynamics inherent to the financial markets. Similarly, collusion amongst construction firms in a number of regions in Japan in the 2000s can be identified with the formation of clusters of anomalous highly connected companies.*

## Analysis

### *Dynamics of financial markets*

In order to investigate the dynamics of financial markets we have developed a simple multi agent network model for a basic financial system, comprising of three fundamental types of agents: Banks, Investors and Borrowers (see Methods section for details). Our approach to modeling this system is inspired by the modeling of societies and ecosystems, in which a key role is played by the virtual intra and interdependence of species [1][2][3][4]. This translates in our model into a focus on: (i) the dynamics of infection of business strategies within the banking sector and of culture dissemination within the investment and fund management community, and (ii) the topological aspects of the network of interactions.

[†] Complexity & Networks Group and Department of Mathematics, Imperial College London, South Kensington Campus, SW7 2AZ, e.viegas11@imperial.ac.uk, h.jensen@imperial.ac.uk.

[††] Department of Computational Intelligence and Systems Science, Tokyo Institute of Technology , 4259-G3-52 Nagatsuta-cho, Midori-ku, Yokohama, 226-8502 Japan, takayasu@dis.titech.ac.jp, miura@smp.dis.titech.ac.jp, tamura@smp.dis.titech.ac.jp.

[§] The Canon Institute for Global Studies, 11F, Shin-Marunouchi, Chiyoda-ku, Tokyo, 100-6511 Japan, ohnishi.takaaki@canon-igs.org

[‡‡] Sony Computer Science Laboratories, 3-14-13 Higashigotanda, Shinagawa-ku, Tokyo, 141-0022 Japan, takayasu@csl.sony.co.jp

[§§] Meiji Institute of Advanced Study of Mathematical Sciences, Meiji University, 1-1-1 Higashimita, Tama-ku, Kawasaki, 214-8571, Japan

In order to focus more clearly on the influence of the collective action of agents, and their interaction amongst themselves on system stability, we built this model in a manner that it does not required detailed financial and economic data as inputs, relying solely on historical interest rates. This is not a limitation of the approach, but a simplification to enable a clearer interpretation of results.

Our model differs from traditional quantitative finance in that it does not focus on risk types (credit, market, liquidity, etc.) or risk quantification (for example Value-at-Risk, Probability of Default, Loss-Given-Default,)$_{(5)}$. Instead, our model makes use of risk parameters purely as a relative measure to rank agents, and to ensure that the relationship between expected losses and expected returns is always consistently maintained. Our approach also differs from traditional behavioral finance models that typically focus on individual behavior as the drivers for decision making$_{(6)}$. The present model contains elements similar to previous work such as Johansen, Ledoit, and Sornette$_{(7)}$ in introducing feedback mechanisms and copies of behaviors. It differs in being concerned with a broader general system dynamics, and in particular the relation between diversity and stability. Our network analysis is also inspired by the work carried out by Inaoka, Ninomiya, Taniguchi, Shimizu and Takayasu$_{(8)}$ on the networks of banking transactions, differing to the extent that we focus on the relationship between bank and investor instead of the inter-banking transactions.

We find typically that the dynamics of a collection of interdependent financial agents leads to strong homogeneity in the longer term, and that this lack of diversity leads to the emergence of unstable periods. In regulatory terms this suggests that the current existing rules in the most developed countries may in effect contribute to the instability of the financial system, by enforcing homogeneity across business models and therefore reducing diversity.

Figure 1 compares the emergence of banking crisis over circa 900 model simulations to two pieces of US economic data: (i) the number of bankruptcies (failures and financial assistances) in the banking sector in the US, (ii) the years of negative GDP growth (contraction) seen in the US economy. This is done for a period from January 1973 to December 2011. As an input, our model uses actual US base rate movements, and computes the number of bankruptcies under adaptive evolutionary dynamics for investors and banks. To isolate the importance of the evolutionary aspects of the dynamics, we also include the results calculated from a purely conventional market dynamics, excluding the evolutionary dynamics.

This figure shows that only when evolutionary dynamics is included in the model, do the results compare well with the sequences of bankruptcies in the US, as well as the periods of economic contraction. We note that for simulations resulting in the emergence of two crises, the results are also significantly in line with the beginning of the two periods of banking crisis[1] in the US, that are normally described in the economic literature: (i) the "Savings and Loans Crisis", which is normally dated from the early 80s to the early 90s[9], and (ii) the more recent "Subprime Crisis" dated 2007[10].

---

[1] We note that it is important to make a distinction between financial crisis and banking crisis. Banking crises are normally associated with failures and bankruptcies of financial institutions within a country, whereas financial crises normally have a broader connotation involving various economic factors (GDP, unemployment, currency, trade, etc.). The Emerging market crisis in the late 90s, for example, had significant impact in the US and other Western Economies, but did not result in a higher number of banking failures within the US.

In contrast, no crisis arises when the evolutionary dynamics are not present which is in line with the expectations based solely on economic theory of rational market equilibrium.

Figure 2 provides us with snapshots of the underlying structure of our simulated banking network for a single model realization under evolutionary dynamics, characterized by crises occurring in 1981 and 2008. For comparative purposes, we also present the results of a separate single realization without evolutionary dynamics.

It is clear that the adaptive behavior of agents, arising from the evolutionary dynamics, leads to the emergence of a dominant strategy, and consequently to a significant reduction in the diversity of bank strategies. Those dynamics also give rise to increased investors' return expectations. These effects are particularly pronounced during the period before the second (and larger) financial crisis.

The financial crises only arise as a result of the evolutionary dynamics, and furthermore the model suggests that the nature of each crisis is more complex than a simple linear relationship between the levels of diversity and the market conditions. Prior to both financial crises, the emergence of a dominant strategy can be observed (see Figure 2B). However, there are marked differences between these crises arising within the model output: (i) As it can be noted in Figure 2, the number of banks with the dominant strategy during the first crisis is significantly smaller than the number of banks with the dominant (and different) strategy in the second crisis; (ii) The dominant strategy accounts for only 30% of all bankrupt agents during the period of the first crisis, compared to a total of 90% for the second crisis. In addition, relative to the other strategies, the dominant strategy for the first crisis is conservative, whereas the dominant strategy for the second crisis is aggressive.

Those results are qualitatively consistent with the fundamental nature of the actual US crises given that the "Loan and Savings Crisis" mostly affected Thrifts (i.e. US financial institutions that accepts saving deposits and invest and invest in mortgages and personal loans) which were supposed to be conservative institutions with varied profiles. In contrast the "Subprime Crisis" resulted in most financial institutions adopting similar strategies, either originating subprime loans or investing in higher yield mortgage bonds that were used to fund those loans.

In addition, Figure 2B indicates that the dominant strategy leading to the second crisis initially started its development during the mid 90s, a date which is fully consistent with the real life beginnings of the subprime mortgage market in the USA[11].[2]

In the real financial world we also observe that reduction of diversity is partially a result of mergers and acquisitions. One striking statistic is the significant reduction of the numbers of financial entities in the USA from 13,976 in 1973 to 6,290 in 2011[3]. In addition, the gradual erosion of regulation separating financial activities, such as the more liberal interpretation and subsequent repudiation of the Glass-Steagall Act, may also have contributed to a higher banking similarity.

---

[2] We observed qualitatively similar behaviour for those runs within similar two crisis model realisations. Whilst outside of the scope of this paper, we plan to expand our work in the future studying in more details the relationship between the crisis, changes to the network dynamics, and the emergence of the dominant strategies.

One point to note is that whereas within the model, many banks with the same strategy remain distinct, in practice this may correspond to mergers and acquisitions among banks. As a result, some agents within the dominant strategy may in the real world form part of a single financial group.

To summarize, our model suggests that a valuable indicator of loss of systemic stability can be obtained from a bank network analysis and that, moreover, attempts to forecast the trajectory of the financial system must take into account the adaptive evolutionary aspects of the financial entities.

### *Japanese firm network*

To further demonstrate the potential of topological aspects of the network analysis as a diagnostic tool, in this Section we expand our analysis into a real world network dataset. Specifically, we analyzed an exhaustive set of data from business dealings of Japanese firms in 2005 provided by Tokyo Shoko Research, Ltd. (TSR). The Japanese inter-firm network consists of about one million companies interconnected through nearly 4 million links, corresponding to declared transactions of goods or services between firms.

We are able to identify collusion between construction firms that took place in certain regions of Japan[12][13][14] through a percolation study of the Japanese firm network. Even though this collusion during the 2000s is now well known, it was not discovered at the time. If such an analysis had been applied back then, the collusion could have been detected as events were unfolding.

We test the stability of the network in two ways. Our first test consists of removing the company with the largest number of connections and with probability *p* removing the neighbors of this company followed by removing the neighbors of the neighbors with probability *p* and so forth. We find that when *p* exceeds the value $p_c$ = 0.012 this process propagates throughout the whole network. From analyzing the Japanese firm bankruptcy data we can evaluate the probability of contagious failure as *p*=0.010, which is lower than the critical value but worryingly close to it to be of concern as the bankruptcy of a major firm might lead to a collapse of the entire economical network.

In our second test we use the standard procedure for complex networks[15] of gradually removing firms in descending order starting with the most connected companies first. By removing about 27% of the most well connected firms the Japanese firm network falls apart losing the connectivity across the whole country. This value is significantly above the corresponding value of 24% for randomized networks having the same distribution of link numbers.

---

[3] Source: Federal Deposit Insurance Corporation – FDIC – Table CB14.

The enhanced strength of the real network is to a large extent a natural and healthy consequence of sector structure and collaboration. However, an inspection of the geographical distribution of the companies remaining in the network at the threshold for disintegration, see Figure 3, reveals inappropriate practices. In Figure 3a an example of companies remaining at the threshold in the case of randomized artificial network is shown, in which companies are distributed widely proportional to the population density. On the other hand in Figure 3b, too many of the surviving companies are construction companies located in a few restricted areas, Wakayama prefecture, Nagoya city and Fukushima prefecture. Our analysis has uncovered the well-known collusion affairs amongst construction firms, which took place in parallel in Wakayama, Nagoya and Fukushima around 2005, see e.g. ((12),(13), and (14)).

## Method

(1) *The Financial Market Network and Dynamics.*

The financial market is represented by three basic agents: Investors, Banks and Borrowers. The interaction among these agents is simulated on a cycle by cycle basis, with each cycle representing a month from January 1973 to December 2011.

We provide in Figure 4 below, a schematic representation of a Bank, and related cash flows for each cycle.

    a) Financial Markets Under "Conventional Dynamics"

In Figure 5, we provide a schematic representation of the methodology we define as "Conventional Dynamics", together with a summary describing those dynamics. For a more detailed description, we refer the reader to Appendix 1 of this document and the formulae therein. For clarity, a list of relevant variables that appear in the model is presented in Appendix 2.

Investors are on the top of the economic structure, providing funding to banks. Each investor is characterized by their investment return expectation $^{in}\text{Rex}(t)$, and their total wealth. The feedback from actual returns interacting with expected returns generates an investor risk appetite parameter $^{in}q(t)$, according to the standard deviation of the downside risk [16].

Based on the relative magnitude of their risk appetite parameter, Investors are then ranked and aggregated into a number of 11 categories of rating preferences. The number or categories aims to replicate the investment grade ratings assigned by international rating agencies, which ranges from AAA to BBB-. An investor's wealth is divided into smaller investment portions, or tranches ($^{in}Tr_n$ - see Appendix for details of notation), based on a pre-defined concentration limit.

Banks are the intermediate agents within the economic structure, capturing cash flows from investors and placing loans to borrowers or within the interbank market.
Banks are characterized by their (i) Capital Amount, (ii) Target Shareholder Return, (iii) Target Capital Ratio, and (iv) Bonus Ratio. For each of the banks, financial data and flows are structured through basic accounting principles as demonstrated in Figure 4. A Bank's strategy is given by the sum of Target Shareholder Return and Bonus Ratio (retained to 4 decimal places).

Banks are ranked and aggregated into 11 rating categories as a function of their Target Capital Ratio and Target Shareholder Return. The Target Capital Ratio also drives the borrowing limits of each bank. Banks go bankrupt when their Actual Capital Ratio is below the regulatory minimum requirement of 8% and require financial assistance when they can no longer capture monies from investors.

Monies are allocated from Investors to Banks through a selection process, which begins by randomly selecting an Investor tranche, $^{in}Tr_n$. Banks are then ordered based on the absolute difference between their rating category and the investor rating preference; the probability that the tranche is invested is then given by a function of the closeness of those ratings.

A similar process is followed in relation to the allocation of monies from Banks to Loans, with the probability of investment derived as a function of the closeness between the price of the Loan and the Banks required return (=Benchmark Return).

The Cost of Borrowing is derived from the investors return expectations, $^{in}Rex(t)$, and the offer and cash demands for each of the rating categories. The Cost of Borrowing, together with the Target Shareholder Return, generate the Banks' required return.

Loans are the representation of borrowers, and it is their demand that drives the circulation in the system. The loans are characterized by a relative performance parameter $^{ls}q$ (where $0 < {}^{ls}q < {}^{ls}q_{max}$). A low value of $^{ls}q(t)$ indicates a low probability of default, while a high value of $^{ls}q$ would signify a mortgage that has a high probability of default. The relative performance parameter drives the probability of default of the loan at redemption, as well as the Price of the Loan at origination.

The parameter $^{ls}q$ is expressed by the cumulative distribution function of a log-normal distribution, as below:

$$^{ls}q(t) = \frac{1}{2} \operatorname{erfc}\left( \frac{\ln\left(5 \frac{\max(ls)+1-ls}{\max(ls)}\right) - \mu(t)}{\sqrt{2\sigma^2}} \right)$$

The parameter $\sigma^2$ represents the variance of the distribution and it is fixed at 0.5, whereas $\mu$ represents the mean of the distribution.

The changes to the parameter $\mu$ is derived from monetary policy, namely the changes to the Base Rate $ir_{(t)}$, where we have used the actual US data, through the following formula:

$$\mu(t) = \ln(ir(t) + 1) * c \text{ , where } c \text{ is a scaling parameter equal to 2.71.}$$

a) Evolutionary Dynamics

We then modify the Bank and Investor agents through the mechanism of Culture Dissemination within the Investors' Community and Infection of the Bank's Business Strategy.

Investors are ordered based on their actual returns achieved over the last 24 months, and the benchmark level is equivalent to that of the investor within the 40th centile from the most to least profitable. Investors below this benchmark adjust their $^{in}$Rex according to the following formula:

$$^{in}Rex(t) = {^{in}Rex(t-1)} + \frac{a}{4}e^{(b*ir(t))}$$

The changes in the investment return expectations $^{in}Rex(t)$ of Investors is modeled through a modified version of the Axelrod model for the dissemination of culture[17][18].

This is applied in the model using a stochastic process in which every investor earning returns below the benchmark has a probability of increasing $^{in}Rex(t)$ towards the benchmark. Performances above benchmark do not result in changes $^{in}Rex(t)$ given that, as described above, it is assumed that investors evaluate risk and returns on the basis of the typical standard deviation of downside risk.

The parameters *a* and *b* are set at 0.02891 and -0.2168, regulating the speed of movement towards benchmark.

The process of less successful banks copying the business strategies of the more successful is inspired by the bacterial conjugation process[19], where the Target Shareholder Returns and Bonus Ratio parameters are copied from the most profitable Bank and replicated into the less successful one if those are higher. The model assumes a uniform probability of infection of the business strategy for all bank agents at 1% per year.

The replication of the $^{bk}$SR(t) and $^{bk}$BN(t) parameters is based on the principle that within the Financial Markets, the performance is fundamentally judged on a bank's return on employed capital, and the ability to pay staff bonuses so that both external and internal stakeholders are satisfied.

(2) *The analysis of the Japanese firm network.*

The disintegration analysis is done by first removing firms one by one in descending order (i.e. the largest sale firm first, then the second largest, etc). The ratio of the number of removed firms over the total number of firms is the control parameter, which we call *f*. After each removal, we calculate *Q*, the size of the largest strongly connected cluster (LSCC) as the order parameter, where "strongly connected cluster" means the set of nodes (=firms) which are connected mutually by some business relation (buy or sell).

The critical point is defined by the value of $f\_c$, which satisfies the relation, $Q$ decreases proportional to a power of ($f\_c - f$).

This is the ordinary percolation phase transition, for $f$ smaller than $f\_c$ the LSCC is large enough that it is spanning over the whole country. For $f$ larger than $f\_c$ the LSCC is very small, namely, the network is totally broken into small pieces. The fact that the value of $f\_c$ for the real network is significantly larger than the randomized case implies that smaller sale firms tend to have business relations with smaller sale firms in the same location. Bid-rigging emerges in the analysis as the extreme limit of this tendency.

The bottom panel in Figure 3 is for the real business network and the top panel refers to an artificial randomized network. In the randomized network, a link connecting a pair of firms is swapped with another randomly chosen link. In more detail, assume that a firm A sends money to a firm B, and a firm C sends money to a firm D, then by swapping the links, money flows from A to D, and C to B. By repeating this random swapping of links millions of times we have a randomized network in which the degree distribution is invariant. In the randomized network the critical point is lower and there remains more firms at the critical point of disintegration reached by removing firms in descending order of sales. Note that in both randomized and the real network the dots represent firms which constitute one connecting cluster by trading interaction.

## References

(1) Haldane, A. G. (2009): "Rethinking Financial Networks", Speech delivered at the Financial Student Association, Amsterdam, April 2009.
(2) Lawson, D., Jensen, H.J. and K. Kaneko, K., *Diversity as a product of interspecial interactions.* J. Theor. Biol., 243, 299-307 (2006).
(3) Jensen, H.J. and Arcaute, E., *Complexity, collective effects, and modeling of ecosystems: formation, function, and stability*. Ann. N.Y. Acad. Sci. **1195**, E19-E26 (2010).
(4) Haldane, A. G. and May, R.M., *Systemic risk in banking ecology*, Nature **469**, 352 (2011).
(5) Varotto, Simone (2011) "Liquidity risk, credit risk, market risk and bank capital", International Journal of Managerial Finance, Vol. 7 Iss: 2, pp.134 – 152.
(6) Tversky, A. and Kahneman, D. (1991): "Loss Aversion and Riskless choice: A reference-dependent model",The Quarterly Journal of Economics (1991) 106 (4): 1039-1061.
(7) Johansen, Anders ; Ledoit, Olivier ; Sornette, Didier (2000): "Crashes as Critical Points" International Journal of Theoretical & Applied Finance Apr2000, Vol. 3 Issue 2, p219.
(8) Inaoka, H.; Ninomiya T., Taniguchi K.; Shimizu, T.; Takayasu H (2004): "Fractal Network derived from banking transaction – An analysis of network structures formed by financial institutions", Bank of Japan Working Paper No.2004-E-04, Bank of Japan, https://www.boj.or.jp/en/research/wps_rev/wps_2004/data/wp04e04.pdf
(9) Federal Deposit Insurance Corporation - FDIC (1997) – "An Examination of the Banking Crisis of the 1980s and Early 1990s". History of the Eighties – Lessons for the Future, Volume 1.

(10) Crouhy, M.G.; Jarrow, R. A. and Turnbull, S. M.(2008) – "The Subprime Credit Crisis of 07", September 12, 2007 revised July 4, 2008.
(11) Chomsisengphet, S.; Pennington-Cross, A. (2006) – "The Evolution of the Subprime Mortgage Market", Federal Reserve Bank of St. Louis *Review*, Jan/Feb 2006, 88(1), pp. 31-56.
(12) Daily Yomiuri 29 Nov 2006. Arrested Wakayama Gov. Yoshiki Kimura reportedly has reversed earlier statements and admitted his involvement in bid-rigging.
http://article.wn.com/view/2006/11/29/Wakayama_governor_admits_to_bidrigging/.
(13) The Japan Times, 24 Jan 2007. Nagoya bid-rigging scandal spreads to highways. http://www.japantimes.co.jp/print/nn20070124a6.html.
(14) The Japan Times, 28 Sep 2006. Governor of Fukushima steps down over brother's bid-rigging arrest.
http://www.japantimes.co.jp/print/nn20060928a6.html
(15) Newman, M. E. "Networks: An Introduction", Oxford Univ. Press (2010).
(16) Developing a Risk Rating Methodology, Joint ABI and IMA Research Paper (2010).
(17) Axelrod, Robert (1997): "The Dissemination of Culture: A Model with Local Convergence and Global Polarization", The Journal of Conflict Resolution, Vol. 41, No. 2 (Apr., 1997), pp. 203-226.
(18) Lanchier, N (2010): "The Axelrod model for the dissemination of culture revisited", Apr 2010.
(19) Lederberg, J. (1960): "A view of Genetics", Science, Volume:131 Issue:3396 Pages:269-276

## Acknowledgements

The authors are grateful to the Research Institute of Economy, Trade and Industry (RIETI) Japan for allowing us to use the TSR data. We also gratefully acknowledge the support from Dr. Robert J. Bozeat on providing an independent critique on the key features of the dynamics of the financial markets, and Dr. Stuart P. Cockburn on the support to the development of the figures within this paper.

The work presented was supported by a grant from the DAIWA foundation.

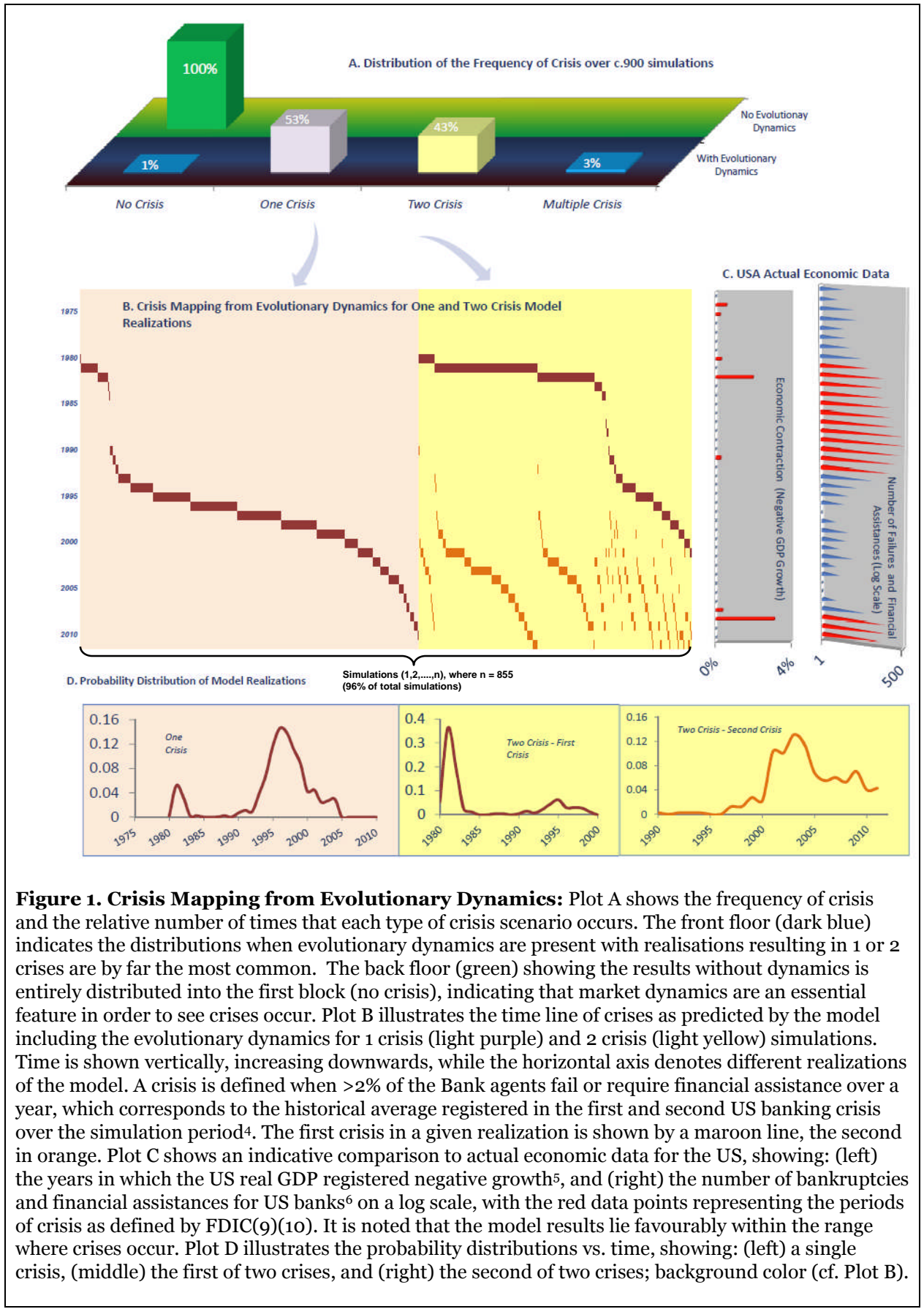

**Figure 1. Crisis Mapping from Evolutionary Dynamics:** Plot A shows the frequency of crisis and the relative number of times that each type of crisis scenario occurs. The front floor (dark blue) indicates the distributions when evolutionary dynamics are present with realisations resulting in 1 or 2 crises are by far the most common. The back floor (green) showing the results without dynamics is entirely distributed into the first block (no crisis), indicating that market dynamics are an essential feature in order to see crises occur. Plot B illustrates the time line of crises as predicted by the model including the evolutionary dynamics for 1 crisis (light purple) and 2 crisis (light yellow) simulations. Time is shown vertically, increasing downwards, while the horizontal axis denotes different realizations of the model. A crisis is defined when >2% of the Bank agents fail or require financial assistance over a year, which corresponds to the historical average registered in the first and second US banking crisis over the simulation period[4]. The first crisis in a given realization is shown by a maroon line, the second in orange. Plot C shows an indicative comparison to actual economic data for the US, showing: (left) the years in which the US real GDP registered negative growth[5], and (right) the number of bankruptcies and financial assistances for US banks[6] on a log scale, with the red data points representing the periods of crisis as defined by FDIC(9)(10). It is noted that the model results lie favourably within the range where crises occur. Plot D illustrates the probability distributions vs. time, showing: (left) a single crisis, (middle) the first of two crises, and (right) the second of two crises; background color (cf. Plot B).

---

[4] As calculated by combining data from FDIC tables BF01 and CB14.
[5] Source: Bureau of Economic Analysis - BEA
[6] Source: FDIC table BF01.

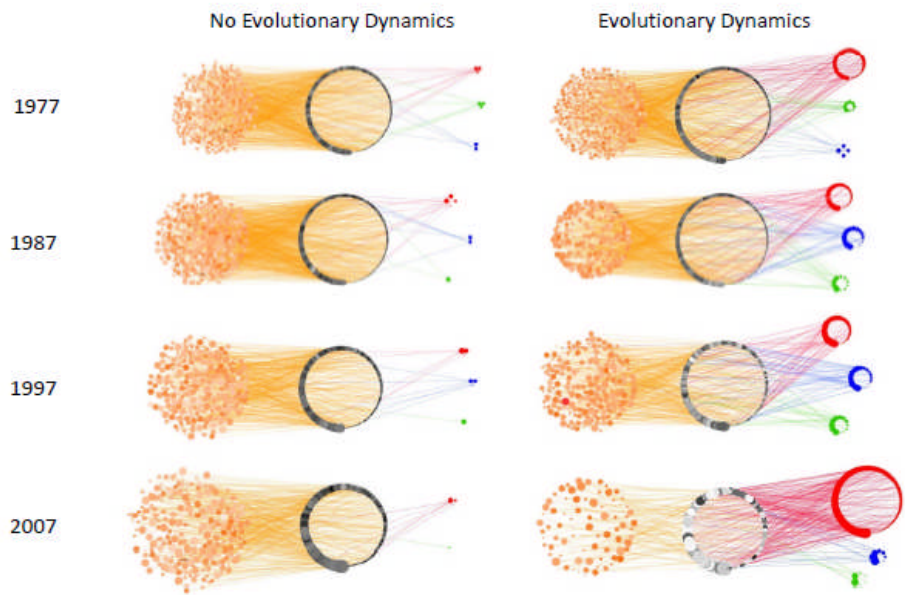
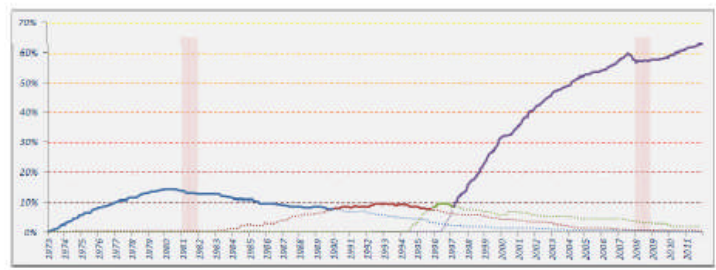

**Figure 2. Effect of evolutionary dynamics upon network evolution:** Plot A shows the comparison of Investor and Bank Strategies at four ten year intervals when evolutionary dynamics is neglected (*left column*) and included (*right column*). For the total network at each time, 1% of constituent entities are selected at random and each entity, i.e. Investors and Banks, is represented by circles, with transactions between these entities denoted by lines. For the Banks, colour is used to represent the strategy, such that the top three strategies are denoted by red, blue and green respectively. The remaining Bank strategies are shown on a scale of light to dark orange. The size of the circles represents the size of the Banks' deposits at each time. Investors are coloured on a black and white scale, indicating a low (black) to high (white) risk appetite. The size of the circles represents Investors' amounts. It is clear that as time progresses, the evolutionary dynamics leads to the emergence of a dominant strategy (large red circle, bottom right plot) and an increase in Investor return expectations. Plot B indicates the evolution of the dominant Bank Strategies. The coloured lines represent each of the strategies that become dominant at a certain stage during the whole model realisation period (horizontal axis). For a given strategy, the line is continuous during the period when that strategy is dominant, and dotted otherwise. The vertical axis represents the number of Bank Agents with a given strategy as a fraction of the total number of active Bank Agents; the shadowed vertical bars mark the years where crises emerge.

**Top Panel**

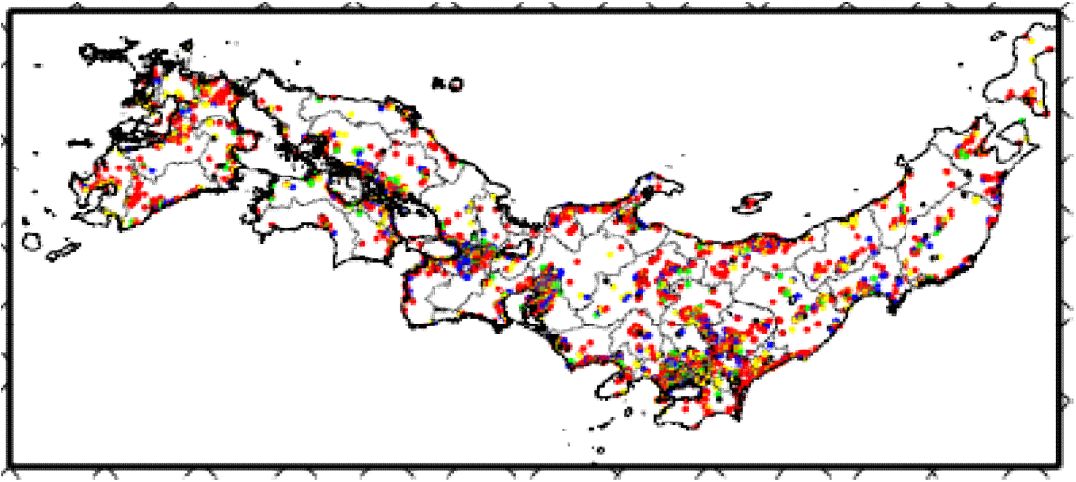

**Bottom Panel**

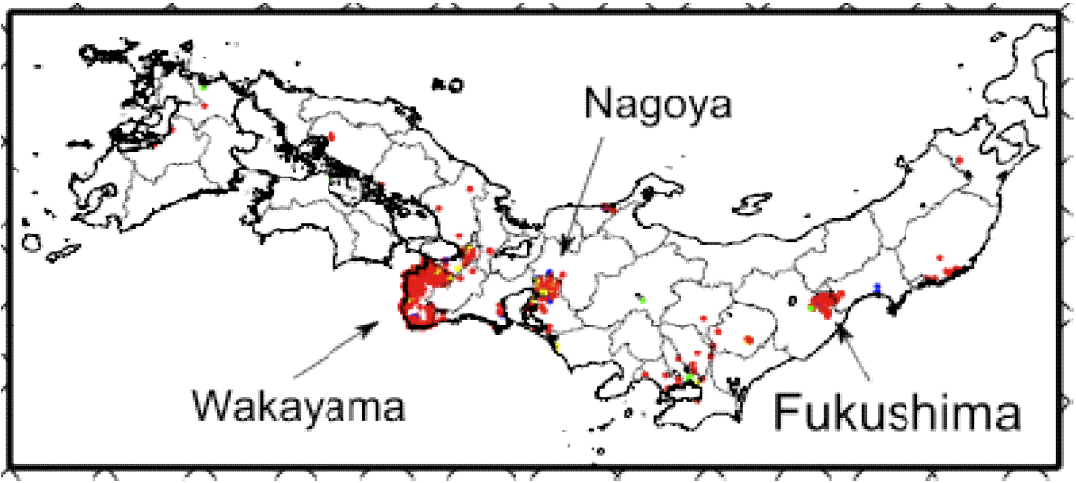

**Figure 3. Japanese Firms Network Mapping:** The top panel shows the distribution of firms (red: construction, blue: manufacturing, yellow: wholesale, green: services, black: others) across Japan at the threshold, as it would be if the interconnectedness were random. The bottom panel shows the actual distribution showing concentration far beyond the level justified by population density in the Wakayama prefecture near the big city Osaka. The analysis reveals that construction firms have excess inter-connections. This is an effect of illegal bid-rigging, which reduces the degree of the firms and leaves them in the network at the critical threshold.

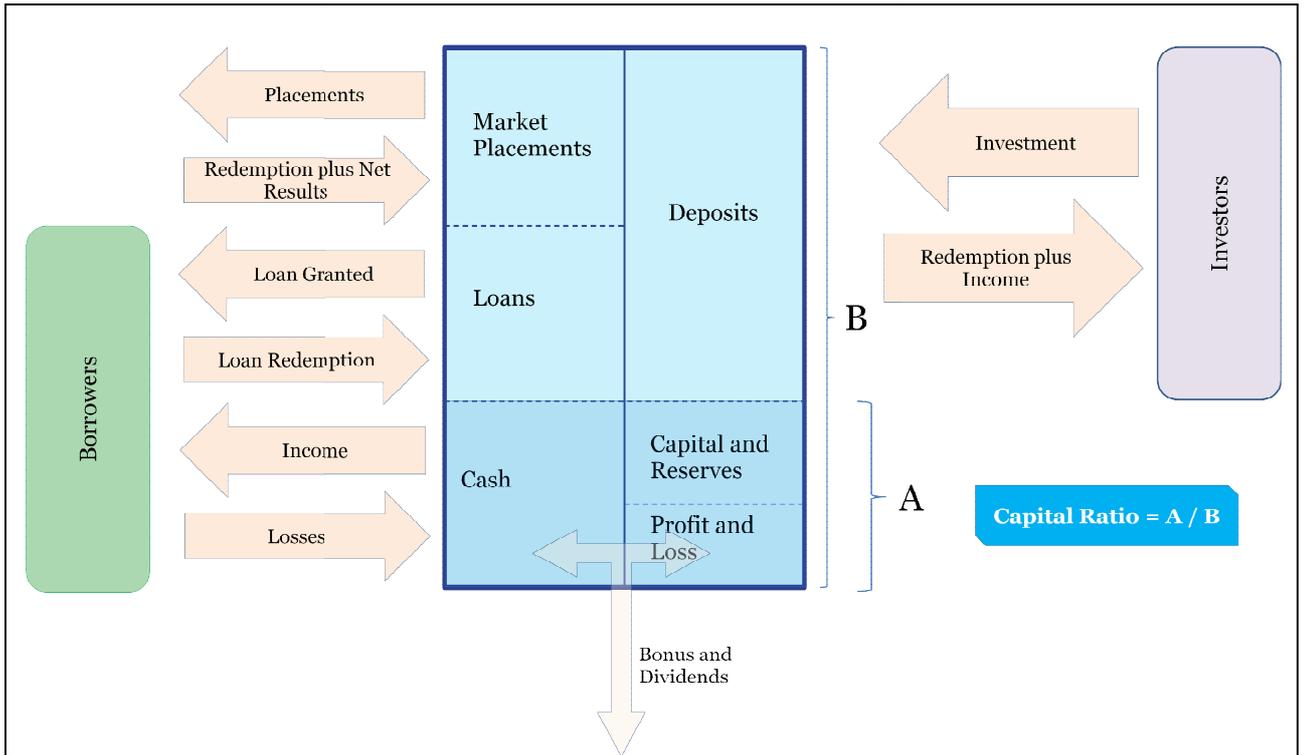

**Figure4. Schematic Representation of a Bank Agent and Money Flows:** Banks capture monies from Investors, and allocate those to Borrowers. Any surplus between Deposits and Lending is considered to be a Market Placement. A Bank's Capital and Reserves are deemed to be retained in Cash or Equivalents. Dividends and Bonuses represent outflows of the Reserves.

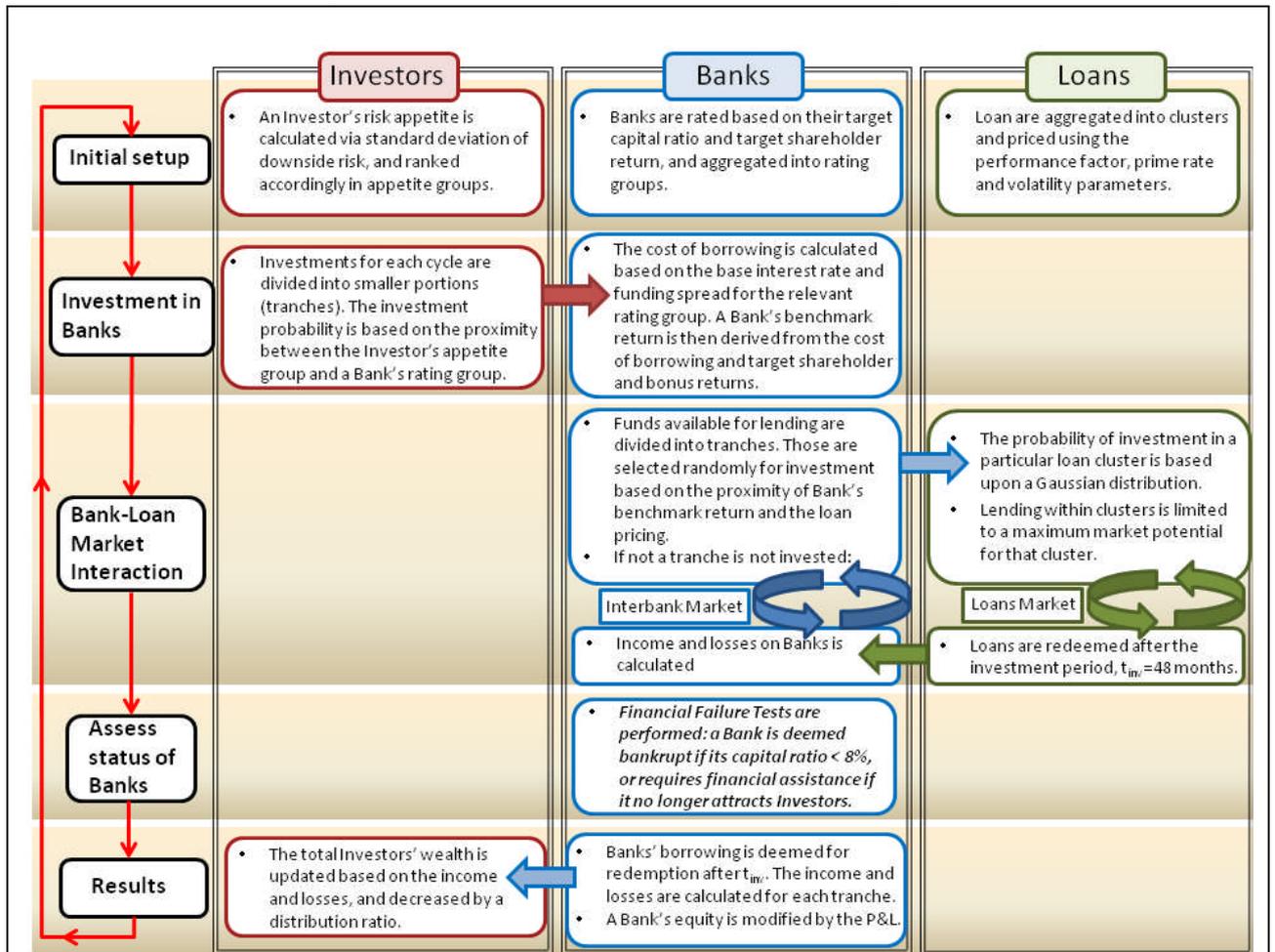

**Figure5. Process Map for "Conventional Dynamics".** The figure shows the key calculations and processes carried out by the model for a given cycle. This excludes the "Evolutionary Dynamics" elements of the full model simulations. See Appendix for a full description and formulae associated to each process.

# Appendices

## *Appendix 1 – Modelling Financial Markets under "Conventional Dynamics"*

To supplement the overview of the methodology for modelling financial markets under "Conventional Dynamics" given in Figure 5, the following table gives a detailed description of this methodology and relevant formulae. As described in the main text, there are three basic market agents: Banks, Investors, and Borrowers, the latter of which are represented below via the behavior of the Loans market.

For convenience, a list of the variables appearing the methodology is presented in Appendix 2, which may prove useful to consult in parallel.

| Entity | Operation | Method (if applicable) |
|---|---|---|
| **Banks** | 1. Rating Assignment<br><br>Banks are ranked as a function of their Target Capital Ratio [$^{bk}TCR(t)$] and assigned a Rating [$^{bk}RT(t)$] ranging from 1 to 10 (with the highest ratio corresponding to a rating of 1).<br><br>$^{bk}RT$ is then adjusted downwards by one notch if the Target Shareholder Return [$^{bk}SR(t)$] is less than the mean of all Banks' Target Shareholder Returns. Therefore 1≤ $^{bk}RT(t)$≤11. | |
| **Investors** | 2. Investors Risk Appetite<br><br>Investors' Risk Appetite [$^{in}q(t)$] is measured through the standard deviation of downside risk.<br><br>We set a period of 24 months as the measurement period, representing the Investor memory period. | $^{in}q(t) = \sqrt{\frac{1}{m}\sum_{t=1}^{m}\left[\min\left(^{in}Ri(t) - {^{in}Rex(t)}, 0\right)\right]^2}$<br><br>where $\begin{cases} ^{in}Rex(t) = \text{Investor investment return expectation} \\ ^{in}Ri(t) = \text{Investors actual return} \\ m = 24 \text{ months} \end{cases}$ |
| **Investors** | 3. Rating Preference<br><br>Investors are ranked as a function of $^{in}q(t)$ and assigned into an Appetite Group [$^{in}AG(t)$] ranging from 1 to 11 (with the lowest $^{in}q(t)$ in group 1). | |
| **Investors** | 4. Investment Tranching<br><br>The available funds to invest, $\Delta^{in}F(t)$, are made through a time period, $t_{inv}$.<br><br>$\Delta^{in}F(t)$ is divided into equal tranches, the $n^{th}$ of which is denoted by $^{in}T_n(t)$, based on the concentration limit ($CL$).<br><br>Within the current run, the concentration limit was set at 10% (resulting in 10 tranches) for each investment period.<br><br>Given that $t_{inv}$ equals 48 months, an | $\Delta^{in}F(t) = {^{in}F(t)} - \sum_{i=t-(t_{inv}-1)}^{t-1}\Delta^{in}F(i)$<br><br>where $\begin{cases} ^{in}F(t) = \text{Total Fund held at given time} \\ t_{inv} = 48 \text{ months} \end{cases}$<br><br>Number of tranches: $nt = \frac{1}{CL}$, where $CL$ equals 10%<br><br>$^{in}T_n(t) = \frac{\Delta^{in}F(t)}{nt}$ |
| | investor total wealth, $^{in}F(t)$, is effectively divided into 480 tranches. | |
| **Banks** | 5. Borrowing Limits<br><br>A bank's borrowing limit [$^{bk}Lim(t)$] corresponds to the maximum amount it can borrow in order to preserve its Target Capital Ratio [$^{bk}TCR(t)$]. | $$^{bk}Lim(t) = {}^{bk}C(t)\left(\frac{1}{{}^{bk}TCR(t)} - 1\right)$$ |
| **Banks** and **Investors** | 6. Cost of Borrowing<br><br>The cost of borrowing for a Bank [$^{bk}CB(t)$] for the funding obtaining at the period ($t$) corresponds to the Base Interest Rate [$ir(t)$] plus Funding Spread [$^{bk}FS(^{bk}RT, t)$] for the Rating [$^{bk}RT(t)$].<br><br>The total amounts deposited from investors into a bank is represented by $^{bk}TD(t)$<br><br>There is a direct mapping between $^{bk}RT(t)$ and $^{in}AG(t)$.<br><br>$^{bk}FS(^{bk}RT,t)$ is calculated for as a function of the demand from banks and offers from within each rating category, and resides between the maximum and minimum investors' expectations in accordance to the adjacent formulae. | $$^{bk}CB(t) = ir(t) + {}^{bk}FS(^{bk}RT, t)$$<br>where:<br>$$\Delta^{bk}Lim(t) = {}^{bk}Lim(t) - {}^{bk}TD(t-1),$$<br>and the Investor Expectation for a given Appetite Group (with $N_{group}$ members) is given by the average of the member expectations, weighted to the funds available to invest:<br>$$Rex_{wa} = \langle {}^{in}Rex({}^{in}q(t) \in {}^{in}AG(t), t) \rangle_w$$<br>$$= \frac{\sum_{i=1}^{N_{group}}({}^{in}Rex_i(t) * \Delta^{in}F_i(t))}{\sum_{i=1}^{N_{group}} \Delta^{in}F_i(t)}.$$<br>To determine the Funding Spread, we use the formula:<br>$$^{bk}FS(^{bk}RT, t) = r(y - z) + w$$<br>and the following conditions,<br><br>if $\sum_{bk} \Delta^{bk}Lim(^{bk}RT = {}^{in}AG, t) < \sum_{i=1}^{N_{group}} \Delta^{in}F_i(t)$ then:<br>$$z = \min[{}^{in}Rex({}^{in}q(t) \in {}^{in}AG(t), t)]$$<br>$$y = Rex_{wa}$$<br>$$w = \min[{}^{in}Rex({}^{in}q(t) \in {}^{in}AG(t), t)]$$<br>and $r = \dfrac{\sum_{bk} \Delta^{bk}Lim(^{bk}RT = {}^{in}AG, t)}{\sum_{i=1}^{N_{group}} \Delta^{in}F_i(t)}$;<br><br>if $\sum_{bk} \Delta^{bk}Lim(^{bk}RT = {}^{in}AG, t) > \sum_{i=1}^{N_{group}} \Delta^{in}F_i(t)$ then:<br>$$z = Rex_{wa}$$<br>$$y = \max[{}^{in}Rex({}^{in}q(t) \in {}^{in}AG(t), t)]$$<br>$$w = \max[{}^{in}Rex({}^{in}q(t) \in {}^{in}AG(t), t)]$$<br>and $r = -\dfrac{\sum_{i=1}^{N_{group}} \Delta^{in}F_i(t)}{\sum_{bk} \Delta^{bk}Lim(^{bk}RT = {}^{in}AG, t)}$ |
| **Banks** and **Investors** | 7. Allocating monies from Investors to Banks<br><br>An investor tranche [$^{in}T_n(t)$] is randomly selected for investment. | If $^{in}T_n(t) < {}^{bk}Lim(t) - {}^{bk}TD(t-1)$ then invest with probability p, |
| | All Banks are then ordered in a priority of investment ranked on a random basis.<br><br>The selected tranche has a probability of investing in a Bank based on its rating equivalence in accordance with the adjacent formulae.<br><br>A similar process in then followed for each subsequent randomly selected tranche.<br><br>$^{bk}TD(t)$ is always capped at $^{bk}Lim(t)$. | where $\begin{cases} \text{if } \lvert {}^{bk}RT(t) - {}^{in}AG(t)\rvert = 0 \rightarrow p = 80\% \\ \text{if } \lvert {}^{bk}RT(t) - {}^{in}AG(t)\rvert = 1 \rightarrow p = 20\% \\ \text{if } \lvert {}^{bk}RT(t) - {}^{in}AG(t)\rvert = 2 \rightarrow p = 10\% \end{cases}$ |
| **Banks** | 8. Total Funding Spread<br><br>A new Total Funding Spread [$^{bk}TFS(t)$] is recalculated for each of the banks. | $^{bk}TFS(t) = \dfrac{\sum_{i=t-(t_{inv}-1)}^{t} {}^{bk}FS\left({}^{bk}RT, i\right) \Delta {}^{bk}TD(i)}{{}^{bk}TD(t)}$<br><br>where<br><br>$\Delta {}^{bk}TD(t) = {}^{bk}TD(t) - {}^{bk}TD(t-1)$ |
| **Banks** | 9. Benchmark Return<br><br>A new benchmark return [$^{bk}BK(t)$] is calculated for each bank. | $^{bk}BK(t) = \dfrac{\dfrac{{}^{bk}C(t) * {}^{bk}SR(t)}{1 - {}^{bk}BN(t)} + {}^{bk}TD(t) * {}^{bk}CB(t)}{{}^{bk}C(t) + {}^{bk}TD(t)} - ir(t)$<br><br>where $\begin{cases} {}^{bk}C(t) = \text{Total Bank's Capital} \\ {}^{bk}SR(t) = \text{Target Shareholder's Return} \\ {}^{bk}BN(t) = \text{Target Bonus Ratio} \end{cases}$ |
| **Banks** | 10. Lending Tranche<br><br>The available funds to lend, $\Delta^{bk}L(t)$, are made through a time period, $t_{inv,}$ equivalent to the same period as the borrowing from investors.<br><br>$\Delta^{bk}L(t)$ is divided into equal tranches, the $n^{th}$ of which is denoted by $^{bk}T_n(t)$, based on the thickness of the tranche ($TT$).<br><br>Within the current run, the tranche thickness was set at 10% (resulting in 10 tranches) for each investment period.<br><br>Given that $t_{inv}$ equals 48 months, an bank's total lending, $^{bk}L(t)$, is effectively divided into 480 tranches. | $\Delta^{bk}L(t) = \Delta^{bk}TD(t)$<br><br>Number of tranches: $nt = \dfrac{1}{TT}$, where $TT$ equals 10% and<br><br>$^{bk}T_n(t) = \dfrac{\Delta^{bk}L(t)}{nt}$ |
| **Loans** | 11. Pricing the Loan<br><br>For each cluster, indexed by $ls$=1, 2,..., 41, Loans are priced as a function of its Relative Performance Parameter ($^{ls}q$), the Prime Rate ($pr$) and a volatility factor ($vol$) in accordance with the adjacent formula. | $^{ls}LP(t) = (pr + {}^{ls}q)(1 + vol)$<br><br>where $\begin{cases} pr = 3\% \\ vol = 20\% \end{cases}$ |
| **Banks** and **Loans** | 12. Allocating monies from Banks to Loans<br><br>A bank tranche ($^{bk}T_n(t)$) is randomly selected for investment. | If $^{bk}T_n(t) < {}^{ls}mrk - {}^{ls}TL(t-1)$ then invest with probability $p$ |
| | All Loan Clusters are then based on the proximity between the bank's benchmark return $^{bk}BK(t)$, and the Pricing of the Loan for the cluster $^{ls}LP(t)$.<br><br>The selected tranche has a probability of investing in a Loan Cluster based in accordance with the adjacent formulae (Gaussian).<br><br>A similar process in then followed for each subsequent randomly selected tranche.<br><br>Total Lending within a cluster [$^{ls}TL(t)$] is capped at the maximum potential market for that cluster [$^{ls}mrk$].<br><br>Tranches that are not allocated to loans are deemed to be placed into the Interbank Market for a period, $t_{inv}$. | $$p = \frac{1}{\sigma\sqrt{2\pi}} e^{-\frac{1}{2}\left(\frac{\left(\left|^{bk}BK(t)-^{ls}LP(t)\right|\right)-\mu}{\sigma}\right)^2}$$<br><br>where $\mu=0$ and $\sigma=1$. |
| **Banks** and **Loans** | 13. Loan Redemption<br><br>Loans are deemed for redemption after $t_{inv}$.<br><br>Income and Losses on Banks are calculated for each invested tranche in accordance with the adjacent formulae. | $$^{bk}Inc(t) = \sum_{1}^{num} {}^{bk}T_n(t) * Rem * \frac{tint}{12}$$<br><br>$$where \begin{cases} \text{if } ^{bk}T_n(t) \text{ Invested in Loans} \\ \text{then } Rem = (^{ls}LP(t-t_{inv}) + ir(t-t_{inv})) \\ \text{if } ^{bk}T_n(t) \text{ invested in Interbank} \\ \text{then } Rem = (Spr + ir(t-t_{inv})) \\ \text{with } Spr = 1\% \end{cases}$$<br><br>$$^{bk}Loss(t) = \sum_{1}^{n}(^{bk}T_n(t) \, ^{ls}q(t)(1-rec)$$<br><br>where $rec = 40\%$, representing recovery levels. |
| **Banks** | 14. Capital Remuneration<br><br>Capital is remunerated according to the adjacent formula. | $$^{bk}Cinc(t) = {}^{bk}C(t)[Spr + ir(t)]/12$$ |
| **Banks** | 15. Bankruptcy Test<br><br>For each bank a bankruptcy test is made. A bank is deemed bankrupt when its Actual Capital Ratio [$^{bk}ACR(t)$] is less than 8% and it requires financial assistance when it can no longer attract Investors funding, so that $^{bk}TD(t)=0$.<br><br>Once a Bank is deemed bankrupt, it is completed removed from future dynamics, whereas a Bank that requires financial assistance might return to operations. | $$^{bk}ACR(t) = \frac{^{bk}C(t)}{^{bk}C(t) + {}^{bk}TD(t)}$$<br><br>$$if \begin{cases} ^{bk}ACR(t) < 8\%, \text{Bank } bk \text{ is bankrupt} \\ ^{bk}ACR(t) > 8\%, \text{Bank } bk \text{ continues trading} \end{cases}$$ |
| **Banks** | 16. Interbank Market Losses<br><br>Interbank deposits are not allocated to other banks on individual basis.<br><br>In the event of a Bankruptcy, Banks within | $$^{bk}IB(t) = {}^{bk}TD(t) - {}^{bk}L(t)$$ |
| | the same Rating category share the losses through allocation on a pari-passu basis according to the formulae adjacent. | $${}^{bk}Lsib(t) = \frac{{}^{bankrupt}IB\left({}^{bankrupt}RT,t\right)}{\sum_{i=1}^{N_{rating}} {}^{bk}IB_i\left({}^{bk}RT,t\right)} {}^{bk}IB(t)$$ |
| **Banks** and **Investors** | 17. Bank's Borrowing Redemption<br><br>Borrowing from Banks are deemed for redemption after $t_{inv}$.<br><br>The cost for the Banks (${}^{bk}Bor(t)$) - and corresponding Income for Investors (${}^{in}Incv(t)$) - and Losses to Investors (${}^{in}Lossv(t)$) are calculated for each invested tranche in accordance with the adjacent formulae. | $${}^{bk}Bor(t) = \sum_{n=1}^{nt} {}^{bk}T_n(t) \; {}^{bk}CB(t-t_{inv}) \; \frac{t_{inv}}{12}$$<br>$${}^{in}Incv(t) = \sum_{n=1}^{nt} {}^{in}T_n(t) * {}^{bk}CB(t-t_{inv}) * \frac{t_{inv}}{12}$$<br>$${}^{in}Lossv(t) = \sum_{n} {}^{in}Tr_n(t)\, \delta_{n,bankrupt}$$<br>where $\delta_{i,j}$ acts as a Kronecker delta, selecting only the bankrupt banks that an Investor invested in. |
| **Banks** | 18. Banks Results and Distribution<br><br>A profit/loss (${}^{bk}NetRes(t)$) for the period for each bank is calculated.<br><br>A Bank's Capital amount is modified by the profit/loss, and decreased by the dividend distribution. | $${}^{bk}NetRes(t) = {}^{bk}Ninc(t)[1 - BN(app,t)]$$<br>where<br>$${}^{bk}Ninc(t) = {}^{bk}Inc(t) - {}^{bk}Loss(t) - {}^{bk}Cinc(t) - {}^{bk}Lsib(t) - {}^{bk}Bor(t).$$<br>$$BN(app,t) = \begin{cases} {}^{bk}BN(t), & {}^{bk}Ninc(t) > 0 \\ 0, & {}^{bk}Ninc(t) < 0 \end{cases}$$<br>$${}^{bk}C(t) = {}^{bk}C(t-1) + {}^{bk}NetRes(t) * (1 - Div(app))$$<br>where<br>$$Div(app) = \begin{cases} 95\%, & {}^{bk}NetRes(t) > 0 \\ 0, & {}^{bk}NetRes(t) < 0 \end{cases}$$ |
| **Investors** | 19. Update of Investors Wealth<br><br>An Investor's wealth is modified by the profit/loss on the investments, and decreased by a distribution ratio. | $${}^{in}F(t) = {}^{in}F(t-1) + {}^{in}Incv(t) - {}^{in}Lossv(t)[1 - Dis(app)]$$<br>where<br>$$Dis(app) = \begin{cases} 90\%, & {}^{in}Incv(t) > {}^{in}Lossv(t) \\ 0, & {}^{in}Incv(t) < {}^{in}Lossv(t) \end{cases}$$ |

## *Appendix 2 – List of Variables*

Each of the three basic agents in the model have variables which represent their essential properties. To make it clear which agent a variable is linked to, we have adopted the following notation:

$$^{Entity}Variable_n(t)$$

For the *Variable* above, which is dependent upon time *t*, *Entity* denotes the agent that the variable is linked to, and takes on the values shown in the Table below.

| Entity | Symbol |
|---|---|
| Investor | in |
| Bank | bk |
| Loans | ls |

The symbols *in, bk ls* also act as labels for the entity types, e.g. *in=1* for Investor number 1, *in=2* for Investor 2, and so on. The subscript index *n* is used to indicate a particular element within a vector quantity. For example, the $n^{th}$ tranche for a Bank is denoted by $^{bk}T_n$; in particular, for Bank number 1 the $n^{th}$ tranche is given by $^1T_n$.

**List of Variables**

**General**
$t_{inv}$: Investment period
$ir(t)$: Base Interest Rate
$Spr$: Spread for Interbank Market

**Banks**
$^{bk}TCR(t)$: Target Capital Ratio
$^{bk}RT(t)$: A Bank's Rating
$^{bk}Lim(t)$: Borrowing Limit
$^{bk}CB(t)$: Cost of Borrowing
$^{bk}FS(^{bk}RT, t)$: Banks' Funding Spread, which is common to Banks with equal $^{bk}RT(t)$.
$^{bk}TD(t)$: Total amount deposited in Banks by Investors
$^{bk}TFS(t)$: Total Funding Spread
$^{bk}BK(t)$: Benchmark Return
$^{bk}C(t)$: Bank's Capital
$^{bk}SR(t)$: Target Shareholder's Return
$^{bk}BN(t)$: Target Bonus Ratio
$\Delta^{bk}L(t)$: Funds available to Lend
$^{bk}T_n(t)$: $n^{th}$ tranche of Bank equity for Loans
$TT$: Tranche Thickness
$\Delta^{bk}L$: Bank's lending in an Investment Period (see $t_{inv}$) to be split across a number of tranches
$^{bk}L$: Bank's total lending, equal to the sum of $\Delta^{bk}L$ over all *m* months, and all tranches within each month. E.g. if m=24 and there are 10 tranches per month, then $^{bk}L$ is split into 240 tranches in total.
$^{bk}Inc(t)$: Bank Income
$^{bk}Loss(t)$: Bank Loss
$^{bk}Lsib(t)$: Loss on Interbank Market
$^{bk}Bor(t)$: Amount paid due to Interest on Borrowing

$^{bk}NetRes(t)$: Profit/Loss
$^{bk}ACR(t)$: Actual Capital Ratio
$^{bk}Cinc(t)$: Remuneration of Capital
$^{bk}CB(t)$: Cost of Borrowing
$^{bk}IB(t)$: Amount invested into the Interbank market

**Investors**
$^{in}q(t)$: Investors' Risk Appetite
$^{in}Rex(t)$: Investor investment return expectation
$^{in}Ri(t)$: Investor's actual return
m: Investor Memory Period
$^{in}AG(t)$: Appetite Group
$^{in}F(t)$: Total fund held at a given time
$\Delta^{in}F(t)$: Available Investor funds for period $t_{inv}$
$^{in}T_n(t)$: Tranches of Investor funds
$CL$: Concentration Limit
$^{in}Incv(t)$: Investor Income
$^{in}Lossv(t)$: Investor Loss

**Loans**
$^{ls}LP(t)$: Loan price for a Loan cluster
$^{ls}q$: Loan Relative Performance parameter
$pr$: Prime Rate
$vol$: Volatility factor
$^{ls}mrk$: Maximum potential market for a Loan cluster